# "First in, last out" solution to the Fermi Paradox

Berezin A.A.

*National Research University of Electronic Technology (MIET)*

**Abstract**

No present observations suggest a technologically advanced extraterrestrial intelligence (ETI) has spread through the galaxy. However, under commonplace assumptions about galactic civilization formation and expansion, this absence of observation is highly unlikely. This improbability constitutes the Fermi Paradox. In this paper, I argue that the Paradox has a trivial solution, requiring no controversial assumptions, which is rarely suggested or discussed. However, that solution would be hard to accept, as it predicts a future for our own civilization that is even worse than extinction.

## 1. Introduction

It is often said that the power of a scientific theory can be calculated as the number of phenomena it explains divided by the number of assumptions it depends on [1]. In that sense, most proposed solutions to the Fermi paradox suffer from severe lack of power. To explain a single phenomenon – the apparent absence of extraterrestrial life in the observable Universe – they often invoke multiple rather controversial assumptions, such as "The Great filters". I believe we can do better.

## 2. Definitions

The cornerstone of the problem is our model of life in the general case. Many proposed solutions take the narrowest definition of Earth-like life and still struggle to come up with a sufficient explanation as to why no life has arisen on any other Earth-like planet, the existence of which seems no longer debated.

However, such a narrow definition is clearly wrong. Even those organisms descending from one common ancestor with ourselves have proven time and time again that we drastically underestimate to what conditions life is able to adapt. And there is no possible way of accounting for all lifeforms that may arise independently throughout the Universe.

Because of that, we have to create a definition that is substrate-invariant. The specific nature of civilizations arising to interstellar level should not matter. They might me biological organisms like ourselves, rogue AIs that rebelled against their creators or distributed planet-scale minds like those described by Stanislaw Lem in "Solaris" [2].

We should, therefore, take a broader definition as the starting point. It has been suggested in [3] to classify as life any objects exhibiting the following traits: homeostasis, organization, metabolism, growth, adaptation, responsiveness and reproduction. For our immediate purposes, this list can be simplified even further.

Homeostasis and internal organization are neither substrate-invariant nor important at the cosmic scale. Metabolism can be generalized as consumption of energy, which is an obvious requirement for any self-organizing system. Adaptation is a consequence of evolution, and since evolution is the only reasonable explanation for complex life regardless of substrate, adaptation should be inherent to it. The same is true for responsiveness to stimuli: even if it is not directly present at the individual level, natural selection is by itself responsive. Growth and reproduction, which are not really different, are the most important for the Fermi paradox because they provide an incentive for life to spread out of its original habitat, and, inevitably, out into space. These properties will be the focus of further discussion.

It should be explicitly noted that we need not invoke intelligence. This saves us one overly complicated and biased definition.

## 3. Parameters

Most proposed solutions to the Fermi paradox have multiple parameters, such as the probabilities of: abiogenesis, multicellular organisms, intelligent life evolving, etc. However, the only variable we can objectively measure is the probability of life becoming detectable from outer space within a certain range from Earth. For simplicity let us call it "parameter A". Depending on whether or not we consider our own civilization detectable, this parameter is either zero or very close to it.

What if an alien civilization appears, but never reaches the stage of space travel or interstellar communication? First, it would be undetectable, and therefore would not solve the paradox. Second, it would have to halt its growth at some point, and no longer fit the definition of life provided earlier. For clarification, I do not suggest that a static civilization is no longer alive, or that we shouldn't treat its individuals morally upon encounter. All I mean is that they are irrelevant to the Fermi paradox. The same reasoning goes for any life stuck on its original planet, be it due to high gravity, unavailability of materials or any other misfortune.

We should, therefore, expect all life in this context to have strong incentives for growth. But to what extent? Obviously, exponential growth cannot be sustained indefinitely. As Isaac Asimov calculated in [4], to continue reproduction at its current rate (at the time), human civilization would have to populate the entire observable universe in just 4200 years. Accounting for the relativistic speed limit, the minimal time limit is 100 thousand years for the galaxy and 500 million for the supercluster [5]. 100 thousand is an insignificant number in evolutionary timelines, considering that it took 3.5 billion years for intelligent life to evolve on Earth. 500 million is considerably more. But again, there's only one significant parameter: how likely it is that several independently arising "lifes" meet in their cosmic expansion phase? This would be parameter B.

We might not know what processes determine the values of A and B, but it is rather obvious that those two sets of processes are nearly identical. Barring the existence of late-stage Great filters, B is just a function of A, and these two variables have a similar order of magnitude. The hypothesis below relies on that assumption.

## 4. Previous explanations

A very similar set of arguments was suggested back in 1981 by Frank Tipler [6]. His interpretation was that extraterrestrial life does not exist, and, therefore, the Fermi paradox is solved. Of course, this was not deemed sufficient explanation by the community at the time. A response [7] came from Carl Sagan and William Newman, pointing out that any intelligent race would make all the same conclusions, then abstain from uncontrolled growth and attempt to destroy any other life that does not impose the same restrictions on itself.

Either idea required further explanation to be considered a solution to the Fermi paradox. In this paper, I am siding with Tipler by adding a crucial detail to his hypothesis.

## 5. Proposal

> *"First in, last out" solution to the Fermi Paradox:* what if the first life that reaches interstellar travel capability necessarily eradicates all competition to fuel its own expansion?

I am not suggesting that a highly developed civilization would consciously wipe out other lifeforms. Most likely, they simply won't notice, the same way a construction crew demolishes an anthill to build real estate because they lack incentive to protect it. And even if the individuals themselves try their best to be cautious, their von Neumann probes [8] probably don't.

This problem is similar to the infamous "Tragedy of the commons". The incentive to grab all available resources is strong, and it only takes one bad actor to ruin the equilibrium, with no possibility to prevent them from appearing at interstellar scale. One rogue AI can potentially populate the entire supercluster with copies of itself, turning every solar system into a supercomputer, and there is no use asking why it would do that. All that matters is that it can.

## 6. Implications

But we are here, our planet and star are still relatively intact, and we are already contemplating the first interstellar probes. Assuming the hypothesis above is correct, what does it mean for our future? The only explanation is the invocation of the *anthropic principle*. We are the first to arrive at the stage. And, most likely, will be the last to leave. The important difference between this proposal and "rare Earth"-type solutions is that human primacy is explained by the anthropic principle alone and not through additional assumptions.

Another interesting implication concerns the predictability of life at large scales. The hypothesis above is invariant of any social, economic or moral progress a civilization might achieve. This would require the existence of forces far stronger than the free will of individuals, which are fundamentally inherent to societies, and inevitably lead it in a direction no single individual would want to pursue. Examples of such forces, such as free market capitalism, are already well-known; however, this hypothesis suggests that resisting them is not nearly as easy as Carl Sagan [7] would like to believe.

But I certainly hope I am wrong. The only way to find out is to continue exploring the Universe and searching for alien life.